\documentclass[aps,pre,twocolumn]{revtex4}

\usepackage{graphicx}
\usepackage{bm}
\usepackage{amsmath}
\usepackage{hyperref}

\begin{document}

\title{Theory of helicoids and skyrmions in confined cholesteric liquid crystals}
\author{Sajedeh Afghah}
\author{Jonathan V. Selinger}
\affiliation{Liquid Crystal Institute, Kent State University, Kent, OH 44242}

\date{February 22, 2017}

\begin{abstract}
Cholesteric liquid crystals experience geometric frustration when they are confined between surfaces with anchoring conditions that are incompatible with the cholesteric twist.  Because of this frustration, they develop complex topological defect structures, which may be helicoids or skyrmions.  We develop a theory for these structures, which extends previous theoretical research by deriving exact solutions for helicoids with the assumption of constant azimuth, calculating numerical solutions for helicoids and skyrmions with varying azimuth, and interpreting the results in terms of competition between terms in the free energy.
\end{abstract}

\maketitle

\section{Introduction}

When a cholesteric liquid crystal is confined between surfaces with homeotropic (perpendicular) anchoring, it experiences \emph{geometric frustration}~\cite{Oswald2005}:  The boundary conditions are incompatible with the favored cholesteric twist.  This geometric frustration is similar to the frustration of a cholesteric liquid crystal in an electric or magnetic field~\cite{Kamien2001}, where the field alignment is incompatible with the cholesteric twist.  Because of this frustration, the confined cholesteric phase forms topological defect structures.  Depending on the geometry and anchoring strength, these defects may be elongated string-like objects called cholesteric fingers or helicoids~\cite{Baudry1998,Baudry1999,Oswald2000,Pirkl2001,Smalyukh2005,Oswald2008}, or they may be localized point-like objects called cholesteric bubbles~\cite{Akahane1976,Bhide1977,Nawa1978,Kerllenevich1981}.

It has recently been recognized that cholesteric bubbles have the remarkable topological properties of skyrmions.  Skyrmions are defects in which the magnitude of the order parameter remains constant, but the orientation varies in a complex texture that cannot anneal away.  Skyrmions were originally proposed in the field of nuclear physics~\cite{Skyrme1962}, and they are now studied extensively in chiral magnets~\cite{Roessler2006,Muehlbauer2009,Yu2010,Lin2013,Banerjee2014}, where they have potential technological applications for magnetic memory and logic.  Hence, cholesteric bubbles or skyrmions are important not only as defects in liquid crystals, but also as examples of the general considerations of geometry and energetics for skyrmions in other physical systems.

Liquid crystal skyrmions have been studied through a range of techniques, including experiments~\cite{Smalyukh2010,Ackerman2012,Chen2013,Ackerman2014,Ackerman2015,Kim2015,Varanytsia2015,Cattaneo2016,Guo2016} and numerical simulations~\cite{Fukuda2010,Fukuda2011PRL,Fukuda2011,Fukuda2012,Guo2016}.  Furthermore, important variational calculations have been done by Leonov et al~\cite{Leonov2014}.  They used a variational theory that was previously developed for skyrmions in chiral magnets, and applied it to chiral liquid crystals.  By minimizing the Frank elastic free energy with appropriate anchoring conditions, they calculated the director texture in both skyrmions and helicoids, and they derived a phase diagram showing the range of parameters in which the system exhibits a skyrmion lattice, a helicoid lattice, or a uniform texture with isolated defects.

The purpose of our current study is to extend the variational calculations of Leonov et al. in several ways.  First, in Sec.~II, we use their free energy with their assumption of a constant azimuthal angle, and derive exact solutions for the director field.   The exact solution can be worked out by conformal mapping for a single helicoid, and by Fourier expansion for a helicoid lattice.  Next, in Sec.~III, we re-examine the assumption of constant azimuthal angle, and show that the three-dimensional (3D) liquid crystal can reduce its free energy by allowing the azimuthal angle to vary.  Without this assumption, we perform numerical calculations of the director field and defect energies for isolated helicoids and helicoid lattices.  In Sec.~IV, we use the same numerical method to investigate isolated skyrmions and skyrmion lattices.  In Finally, in Sec.~V, we compare the free energies to predict a phase diagram for the defect structures.  We interpret the results in terms of a competition among the chiral elastic free energy that favors twist, the non-chiral elastic free energy that penalizes director variations, and the free energy cost of surface singularities.

\section{Helicoids: Exact solutions with assumption of constant azimuth}

Consider a cholesteric liquid crystal confined between two surfaces at $z=\pm d/2$.  In the interior, the liquid crystal director field $\bm{\hat n}(\bm{r})$ has the Frank free energy density
\begin{align}
f=&\frac{1}{2}K_1 (\nabla\cdot\bm{\hat n})^2 +\frac{1}{2}K_2 (\bm{\hat n}\cdot\nabla\times\bm{\hat n})^2 \nonumber\\
&+\frac{1}{2}K_3 |\bm{\hat n}\times\nabla\times\bm{\hat n}|^2 +K_2 q_0 \bm{\hat n}\cdot\nabla\times\bm{\hat n},
\label{frank}
\end{align}
where $q_0$ is the natural twist of the cholesteric phase, and $K_1$, $K_2$, and $K_3$ are the Frank constants for splay, twist, and bend, respectively.
Suppose that the surfaces have strong homeotropic anchoring, which gives the constraint
\begin{equation}
\bm{\hat n}(x,y,z=\pm d/2)=\bm{\hat z}\text{ or }-\bm{\hat z}.
\label{bcn}
\end{equation}

If there were no surface anchoring, the director field would form a cholesteric helix, with the director depending only on one coordinate, which we can call $x$.  In terms of the polar angle $\theta$ and azimuthal angle $\phi$, this helix can be written as
\begin{equation}
\bm{\hat n}(\bm{r})=(\sin\theta(\bm{r})\cos\phi(\bm{r}),\sin\theta(\bm{r})\sin\phi(\bm{r}),\cos\theta(\bm{r})),
\end{equation}
with
\begin{align}
\theta(\bm{r})&=q_0 x,\nonumber\\
\phi(\bm{r})&=-\frac{\pi}{2}.
\end{align}
In the presence of surface anchoring, the director field must be distorted, with a dependence on $z$, in order to satisfy the boundary condition of Eq.~(\ref{bcn}).  In Secs.~II and III, we consider the case of helicoids or cholesteric fingers, where $\bm{\hat n}$ depends on $x$ and $z$ but is independent of $y$.  In Sec.~IV, we consider the case of skyrmions or cholesteric bubbles, where $\bm{\hat n}$ depends on all three coordinates $x$, $y$, and $z$.

To calculate the structure and energy of a helicoid, we must minimize the Frank free energy subject to the boundary condition.  For this calculation, we make the usual assumption of equal Frank constants, $K = K_1 = K_2 = K_3$.  In terms of $\theta$ and $\phi$, the Frank free energy density becomes
\begin{align}
f=&\frac{1}{2}K\biggl[\left(\frac{\partial\theta}{\partial x}\right)^2 + \left(\frac{\partial\theta}{\partial z}\right)^2
+\sin^2 \theta \biggl(\left(\frac{\partial\phi}{\partial x}\right)^2 + \left(\frac{\partial\phi}{\partial z}\right)^2 \biggr)\nonumber\\
\label{generalfreeenergy}
&+2\sin^2 \theta \sin\phi \left(\frac{\partial\theta}{\partial z}\frac{\partial\phi}{\partial x}
-\frac{\partial\theta}{\partial x}\frac{\partial\phi}{\partial z}\right)\\
&+2q_0 \sin\phi \frac{\partial\theta}{\partial x} -2q_0\sin^2 \theta \frac{\partial\phi}{\partial z}
+q_0 \sin2\theta\cos\phi \frac{\partial\phi}{\partial x} \biggr],\nonumber
\end{align}
and the boundary condition becomes
\begin{equation}
\theta(x,y,z=\pm d/2)=0 \pmod\pi,
\label{bctheta}
\end{equation}
with no boundary condition on $\phi$.

With this form of the Frank free energy density, it is not obvious whether the azimuthal angle $\phi$ should be constant with respect to $x$ and $z$.  If $\phi$ were constant, its optimal value would be $\phi=-\pi/2$, in order to minimize the term $K q_0 \sin\phi (\partial\theta/\partial x)$ in the free energy.  In that case, the director distortion would be mainly twist, rather than splay or bend, which should be favorable.  Based on these considerations, Ref.~\cite{Leonov2014} made the assumption that $\phi=-\pi/2$ throughout the cell.  In this section, we make the same assumption, because it allows some exact calculations.  In Secs.~III and IV, we do numerical calculations \emph{without} that assumption on $\phi$.

Using the assumption of constant azimuth $\phi=-\pi/2$, the Frank free energy density simplifies greatly to
\begin{equation}
f=\frac{1}{2}K\biggl[\left(\frac{\partial\theta}{\partial x}\right)^2 + \left(\frac{\partial\theta}{\partial z}\right)^2
-2q_0 \frac{\partial\theta}{\partial x} \biggr],
\label{helicoidfreeenergy}
\end{equation}
and the corresponding Euler-Lagrange equation becomes
\begin{equation}
\frac{\partial^2 \theta}{\partial x^2}+ \frac{\partial^2 \theta}{\partial z^2}=0,
\label{laplace}
\end{equation}
which is just Laplace's equation for the polar angle $\theta$.  This equation can be solved for a single helicoid or for a periodic lattice of helicoids.

\subsection{Conformal mapping for single helicoid}

\begin{figure}
\includegraphics[width=3.375in]{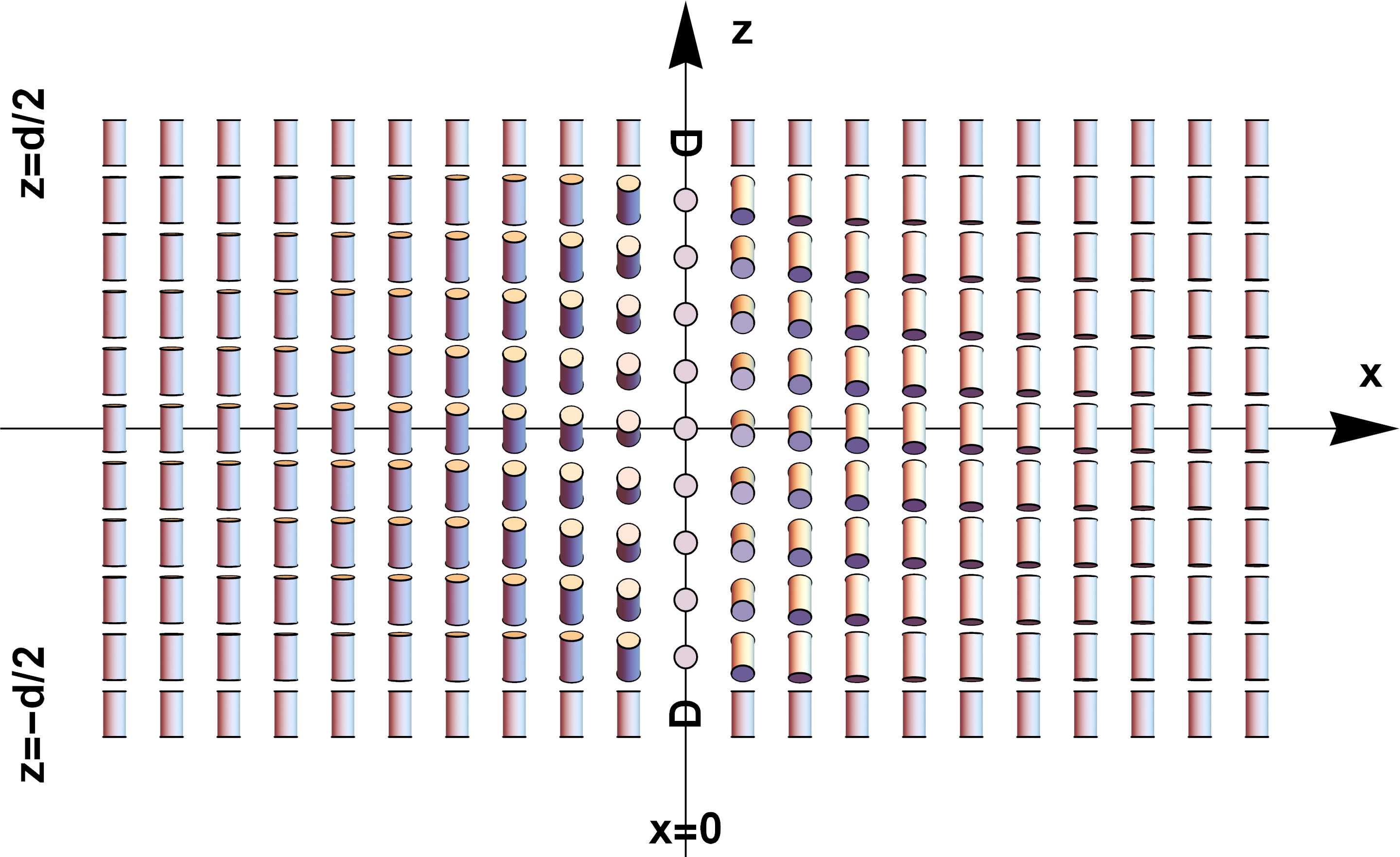}
\caption{(Color online) Single helicoid in the director field of a cholesteric liquid crystal at $x=0$, calculated with the assumption of constant azimuth.  The picture shows the cross section at $y=0$; the director field is extended uniformly forward and backward in $y$.  The symbol D represents disclinations, which are also extended uniformly forward and backward in $y$.}
\end{figure}

To describe a single helicoid, we use the geometry shown in Fig.~1.  Across the helicoid, the director field twists and the polar angle $\theta$ advances by an angle of $\pi$.  Hence, the boundary condition of Eq.~(\ref{bctheta}) becomes more specifically
\begin{equation}
\theta(x,y,z=\pm d/2)=
\begin{cases}
0 &\text{for }x<0,\\
\pi &\text{for }x>0.
\end{cases}
\label{bcsinglehelicoid}
\end{equation}

We can solve Laplace's equation with the boundary condition of Eq.~(\ref{bcsinglehelicoid}) using the method of conformal mapping.  The exact solution is
\begin{align}
\theta(x,y,z)= & \pi+\tan^{-1}\left(\frac{\sin(\pi z/d)-\exp(-\pi x/d)}{\cos(\pi z/d)}\right)\nonumber\\
&-\tan^{-1}\left(\frac{\sin(\pi z/d)+\exp(-\pi x/d)}{\cos(\pi z/d)}\right).
\label{conformal}
\end{align}
This result can be verified by explicit substitution into the differential equation and boundary condition.  It is illustrated by the director field in Fig.~1.

Note that this solution has a characteristic width in $x$ that depends on the height $z$.  It is widest at the midplane $z=0$, where the width is of order $d$.  The width becomes narrower as $z$ approaches the top and bottom surfaces, and it goes to zero right at the surfaces.  Hence, the $\theta$ variation becomes concentrated in a pair of surface disclinations at $x=0$ and $z=\pm d/2$.  These disclinations are lines that run along the surfaces in the $y$ direction.  At the disclinations, the director field itself becomes undefined, as indicated by the D symbols in the figure.

To calculate the total free energy of a single helicoid, we integrate the Frank free energy density
\begin{equation}
F_\text{helicoid}=\int_{-\infty}^{\infty}dx \int_{-L_y /2}^{L_y /2}dy \int_{-d/2}^{d/2}dz f, 
\end{equation}
where $\theta$ is given by Eq.~(\ref{conformal}) and $L_y$ is the system length in the $y$ direction.  This integral is logarithmically divergent because of the disclination lines.  To regularize the divergence in a physical way, we note that the free energy density can never exceed a maximum value $f_\text{max}$, which is the free energy density for melting the cholesteric phase into the isotropic phase.  Physically, this maximum free energy density is related to the disclination core radius $a$ by $f_\text{max}\approx K/a^2$.  By imposing $f_\text{max}$ as a hard cutoff on $f$, we calculate the integral numerically to obtain
\begin{align}
F_\text{helicoid}=&\pi K L_y \left[\frac{1}{2}\log\left(\frac{f_\text{max}d^2}{K}\right) + 0.4 - d q_0\right]\nonumber\\
=&\pi K L_y d (q_H - q_0),
\label{fsinglehelicoid}
\end{align}
where
\begin{equation}
q_H = \frac{\frac{1}{2}\log(f_\text{max}d^2 /K) + 0.4}{d} \approx \frac{\log(d/a) + 0.4}{d}
\end{equation}
is the critical twist for a helicoid.  With the typical values of $a\approx 10$~nm and $d\approx 1$~$\mu$m, we obtain $q_H\approx 5$~$\mu$m$^{-1}$.

From this result, we see that the free energy of a helicoid might be positive or negative, depending on the natural twist $q_0$ compared with the critical twist $q_H$:

\emph{Case 1:}  If $q_0 < q_H$, then the helicoid has a positive free energy.  By comparison, a uniform vertical state with $\bm{\hat n}=\bm{\hat z}$ everywhere has zero free energy.  Hence, a helicoid has a higher free energy than a uniform vertical state, and we would not expect to see any helicoids in thermal equilibrium.  Of course, some scattered helicoids may still occur as metastable defects on the uniform ground state.

\emph{Case 2:}  If $q_0 > q_H$, then the helicoid has a negative free energy compared with the uniform state.  In that case, we would expect to see many helicoids in thermal equilibrium.  The equilibrium density of helicoids depends on the interaction between neighboring helicoids.  To determine this density, we must consider a periodic lattice of helicoids in the calculation below.

\subsection{Fourier series for helicoid lattice}

We now consider a periodic lattice of parallel helicoids, running along the $y$ direction, with a spacing of $\lambda$ in the $x$ direction.  In particular, suppose the centers of the helicoids are located at $x=(m+\frac{1}{2})\lambda$, where $m$ is any integer.  To calculate the director field of the helicoid lattice, we can consider one unit cell between the helicoids centered at $x=-\frac{1}{2}\lambda$ and $x=+\frac{1}{2}\lambda$.  The rest of the director field can then be found by repeating the unit cell periodically.

For this calculation, we must solve Laplace's equation~(\ref{laplace}) in the rectangular domain $-\frac{1}{2}\lambda\le x\le\frac{1}{2}\lambda$ and $-\frac{1}{2}d \le z \le \frac{1}{2}d$.  The boundary conditions are
\begin{equation}
\theta(x,y,z)=
\begin{cases}
-\pi/2 &\text{for }x=-\lambda /2,\\
0 &\text{for }z=\pm d/2 ,\\
+\pi/2 &\text{for }x=+\lambda /2 .
\end{cases}
\label{bchelicoidlattice}
\end{equation}
These boundary conditions imply that the director field rotates through an angle of $\pi$ across the unit cell of the structure.  They also require that the director field has disclinations at the corners of the unit cell, and hence at all $x=(m+\frac{1}{2})\lambda$ and $z=\pm d/2$.

In this geometry, Laplace's equation can be solved by separation of variables.  A general solution with the correct symmetry is the Fourier series
\begin{equation}
\theta(x,y,z)=\sum_k A_k \sinh kx \cos kz ,
\label{thetaseries}
\end{equation}
where $k$ is a separation constant with dimensions of wavevector.  The boundary conditions at $z=\pm d/2$ require that $k=j\pi/d$, where $j$ is any \emph{odd} integer.  The boundary conditions at $x=\pm \lambda /2$ then require that
\begin{equation}
A_k = \frac{2\pi\sin\frac{1}{2}kd}{kd\sinh\frac{1}{2}k\lambda}.
\label{fouriercoefficient}
\end{equation}

To calculate the free energy of the helicoid lattice, we insert the solution of Eqs.~(\ref{thetaseries}--\ref{fouriercoefficient}) into the free energy of Eq.~(\ref{helicoidfreeenergy}).  We integrate over the unit cell, and divide by the volume of unit cell, to obtain the average free energy per volume
\begin{equation}
\frac{F_\text{helicoid lattice}}{\lambda L_y d}=\frac{1}{\lambda d}\int_{-\lambda /2}^{\lambda /2}dx \int_{-d/2}^{d/2}dz f. 
\end{equation}
This calculation can be done exactly for each term in the Fourier series, and the result is
\begin{equation}
\frac{F_\text{helicoid lattice}}{\lambda L_y d}=\frac{\pi K}{\lambda d}\left[\left(\sum_{j\text{ odd}}\frac{2}{j}\coth\frac{j\pi \lambda}{2d}\right)-q_0 d\right]
\end{equation}
For large $\lambda$, this function can be approximated as
\begin{equation}
\frac{F_\text{helicoid lattice}}{\lambda L_y d}=\frac{\pi K}{\lambda d}\left[\left(\sum_{j\text{ odd}}\frac{2}{j}\right)+4 e^{-\pi \lambda /d}-q_0 d\right]
\label{flatticeapprox}
\end{equation}
The summation in Eq.~(\ref{flatticeapprox}) is logarithmically divergent.  Physically, the reason for this divergence is that the free energy includes an integral over the disclinations in the director field at the corners of the unit cell.  To regularize this divergence, we can cut off the sum at a maximum wavevector $k_\text{max}$, which is related to the disclination core radius $a$ by $k_\text{max}\approx\pi/a$, and hence at $j_\text{max}=k_\text{max} d/\pi\approx d/a$.  From the properties of harmonic numbers $H_n$~\cite{MathWorldHarmonicNumbers}, we have
\begin{equation}
\sum_{j\text{ odd}}^{j_\text{max}}\frac{2}{j}=H_{(j_\text{max}/2)}+\log4\approx\log(2j_\text{max})+\gamma,
\end{equation}
in the limit of large $j_\text{max}$, where $\gamma\approx0.577$ is the Euler-Mascheroni constant.  Hence, the free energy of the helicoid lattice becomes
\begin{equation}
\frac{F_\text{helicoid lattice}}{\lambda L_y d}=\frac{\pi K}{\lambda}(q_H - q_0)+\frac{4\pi K}{\lambda d}e^{-\pi \lambda /d},
\label{flatticefinal}
\end{equation}
where
\begin{equation}
q_H = \frac{\log(2j_\text{max}) + \gamma}{d} \approx \frac{\log(d/a) + 1.3}{d}
\end{equation}
With the numerical estimates $a\sim10$~nm and $d\sim1$~$\mu$m, this calculation gives $q_H\approx 6$~$\mu$m$^{-1}$.

The first term of the helicoid lattice free energy~(\ref{flatticefinal}) is equivalent to the single helicoid free energy~(\ref{fsinglehelicoid}), divided by the unit cell volume $\lambda L_y d$.  There is a slight numerical difference in the estimates of $q_H$, which occurs because the disclination cores are treated somewhat differently here than in Sec.~II(A), but that is not important because neither theory gives a precise description of the cores.  More importantly, the lattice free energy~(\ref{flatticefinal}) has a new exponential term $e^{-\pi \lambda /d}$, which shows the extra free energy associated with a helicoid lattice.  It can be interpreted as a repulsive interaction between neighboring helicoids.  It decays exponentially with a decay length of $d/\pi$, proportional to the cell thickness $d$.

\begin{figure}
\includegraphics[width=3.375in]{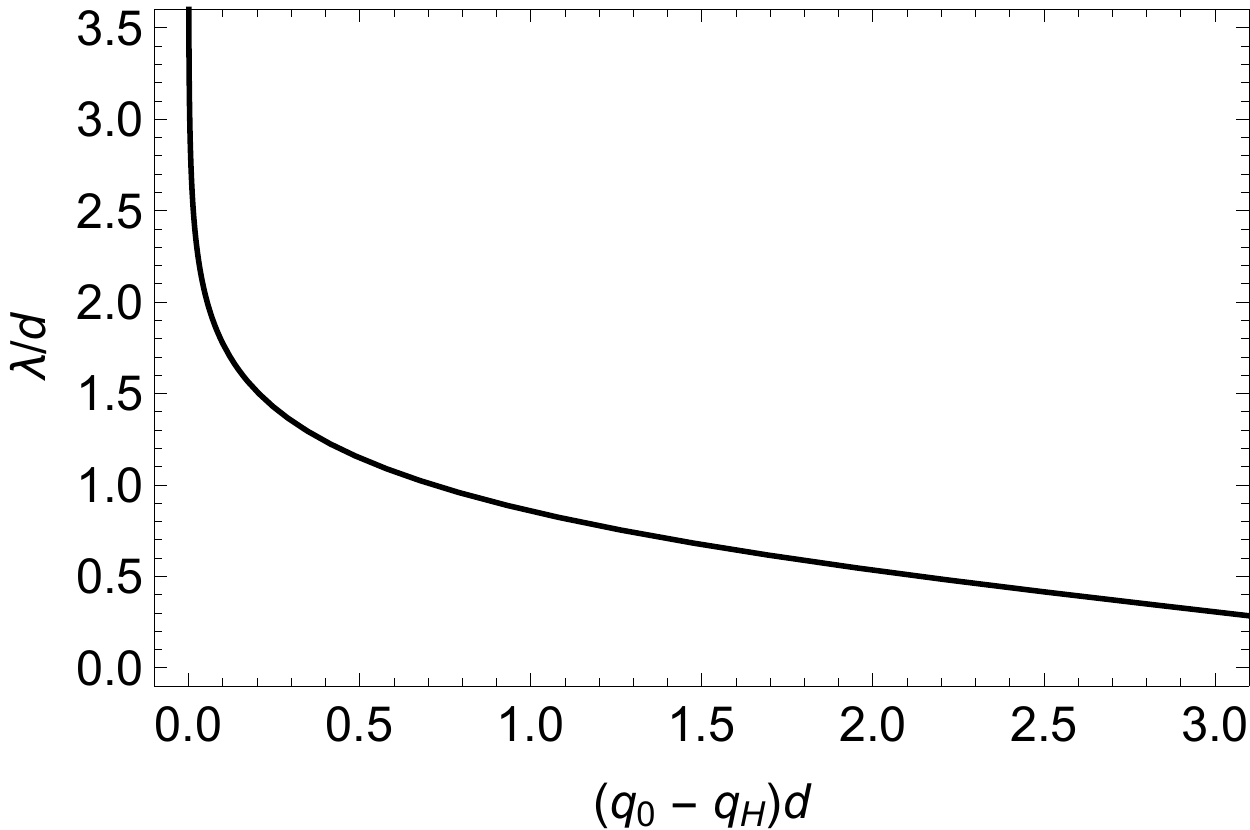}
\caption{Prediction for the helicoid lattice spacing $\lambda$ (scaled by the cell thickness $d$), as a function of the natural cholesteric twist $q_0$ above the critical twist $q_H$ (also scaled by $d$).}
\end{figure}

We can now minimize the average free energy density of Eq.~(\ref{flatticefinal}) to obtain the optimum spacing $\lambda$ between the helicoids.  If $q_0 < q_H$, this calculation gives $\lambda \to\infty$.  In this case, because each single helicoid is unfavorable compared with a uniform state, the density of helicoids goes to zero; i.e.\ helicoids are not present in thermal equilibrium.  By contrast, if $q_0 > q_H$, the minimization gives
\begin{equation}
4\left(1+\frac{\pi \lambda}{d}\right)e^{-\pi \lambda /d}=(q_0 - q_H)d.
\end{equation}
The solution of this equation is shown in Fig.~2.  As the natural twist $q_0$ increases beyond the critical value $q_H$, the helicoid spacing $\lambda$ decreases from infinity.  Over a wide range of $(q_0 - q_H)$, $\lambda$ is close to the cell thickness $d$.

\begin{figure}
\includegraphics[width=3.375in]{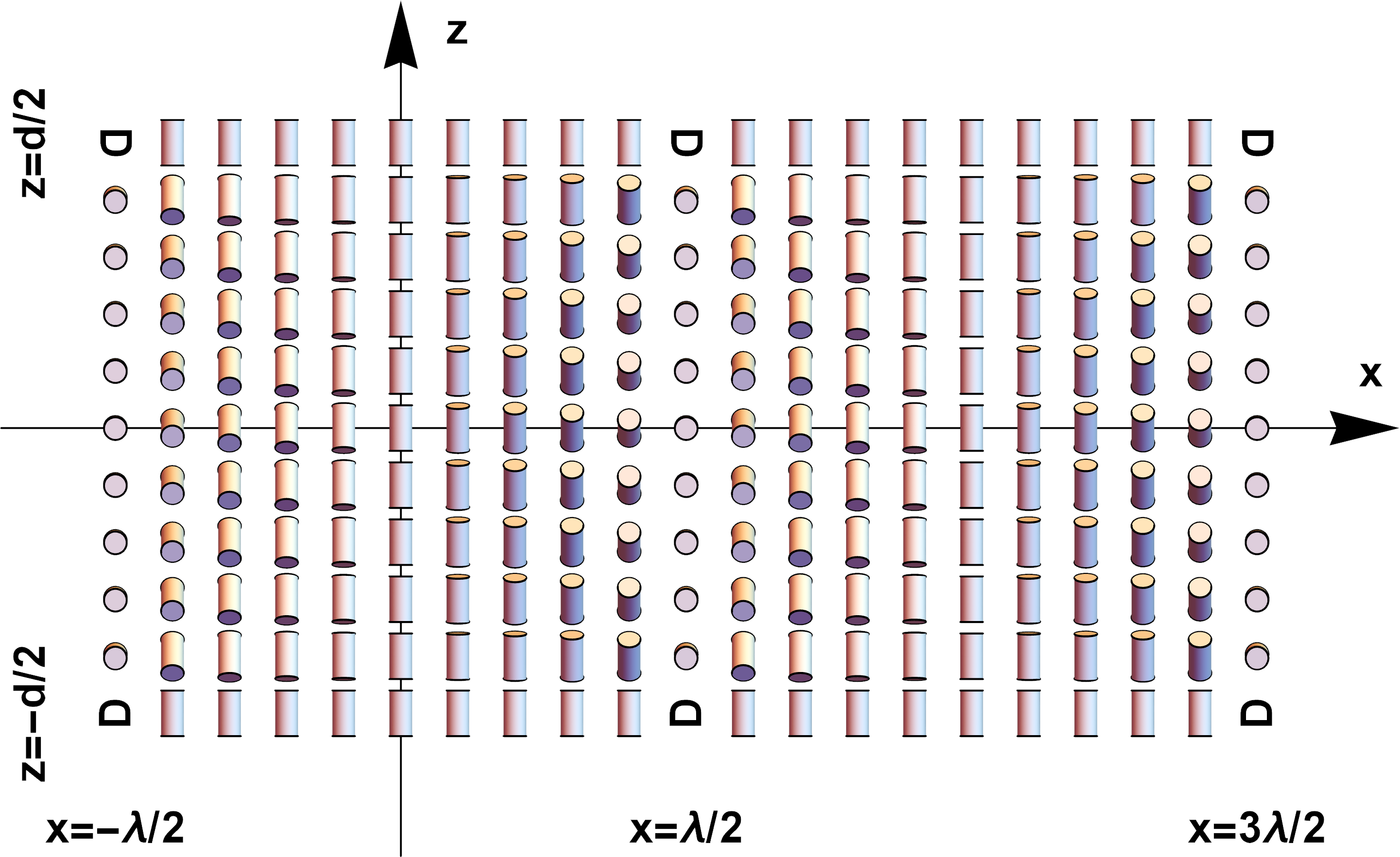}
\caption{(Color online) Lattice of helicoids at $x=-\frac{1}{2}\lambda$, $\frac{1}{2}\lambda$, $\frac{3}{2}\lambda$, \ldots, in the case where $\lambda=d$.  The picture shows the cross sections at $y=0$; the director field is extended uniformly forward and backward in $y$.}
\end{figure}

As a specific example, Fig.~3 shows the director field associated with the series solution of Eq.~(\ref{thetaseries}) when ${(q_0 - q_H)}d = 0.716$, which corresponds to $\lambda/d = 1$.  In the middle of the cell, the director field has an almost uniform twist, quite similar to an unperturbed cholesteric liquid crystal.  Away from the midplane, the twist becomes more concentrated in the helicoids, and the rest of the director field becomes more uniform and vertical.  At the top and bottom surfaces, the twist is localized in the disclinations.

\section{Helicoids: Numerical solutions without assumption of constant azimuth}

We must now re-examine the assumption of constant azimuthal angle $\phi=-\pi/2$, which was made in Ref.~\cite{Leonov2014} and in Sec.~II.

Physically, $\phi=-\pi/2$ would be the optimum angle for twist \emph{if} the director field depended only on $x$.  However, we have already seen that the director field depends on $z$ as well as $x$.  Hence, the cholesteric liquid crystal might be able to reduce its free energy further by varying $\phi$, so that the director field can twist as a function of $z$, in addition to twisting as a function of $x$.

Mathematically, we can calculate the functional derivative $\delta F/\delta\phi(x,z)$ of the integrated free energy from Eq.~(\ref{generalfreeenergy}) with respect to $\phi(x,z)$.  This functional derivative is explicitly nonzero when $\phi=-\pi/2$ and $\theta(x,z)$ is given by Eq.~(\ref{conformal}) for a single helicoid or Eq.~(\ref{thetaseries}) for a helicoid lattice.  Hence, the director field with constant $\phi=-\pi/2$ cannot be the exact minimum of the free energy.

To go beyond the approximation of constant $\phi$, we must minimize the free energy numerically.  For this numerical calculation, we use an algorithm based on relaxational dynamics.  We set up the dynamic equations
\begin{align}
\frac{\partial\theta(x,z,t)}{\partial t}=&-\Gamma_\theta \frac{\delta F}{\delta\theta(x,z,t)},\nonumber\\
\frac{\partial\phi(x,z,t)}{\partial t}=&-\Gamma_\phi \frac{\delta F}{\delta\phi(x,z,t)},
\end{align}
and integrate them forward in time until they converge on a free energy minimum.  The specific choice of dynamic constants $\Gamma_\theta$ and $\Gamma_\phi$ is not important; we set them equal to each other and choose units of time so that $\Gamma_\theta=\Gamma_\phi=1$.  When solving the dynamic equations, we use initial conditions and boundary conditions appropriate for the specific geometry of a single helicoid or a helicoid lattice.

\subsection{Single helicoid}

To model a single helicoid at $x=0$, we solve the dynamic equations on one side of the helicoid, for $0\le x\le L_x$ and $-\frac{1}{2}d\le z\le\frac{1}{2}d$.  Here, $L_x$ is an arbitrary cutoff far from the helicoid, so that the director field is effectively vertical there; we use $L_x = 5d$.  On the other side of the helicoid, for $x\le 0$, the director field can be found by the symmetry $\theta(-x,z,t)=\pi-\theta(x,z,t)$ and $\phi(-x,z,t)=\phi(x,z,t)$.

For the boundary conditions, we require
\begin{align}
&\theta(0,z,t)=\frac{\pi}{2},  &&\phi(0,z,t)=-\frac{\pi}{2},\nonumber\\
&\theta(L_x,z,t)=\pi,          &&\phi(L_x,z,t)=-\frac{\pi}{2},\nonumber\\
&\theta(x,\frac{d}{2},t)=\pi,  &&\phi(x,\frac{d}{2},t)=-\frac{\pi}{2},\nonumber\\
&\theta(x,-\frac{d}{2},t)=\pi, &&\phi(x,-\frac{d}{2},t)=-\frac{\pi}{2}.
\end{align}
The last three boundary conditions on $\phi$ are not important, because the boundary conditions on $\theta$ require that the director field is vertical and hence $\phi$ is irrelevant along the right, top, and bottom boundaries, but we include these boundary conditions as part of the numerical algorithm.  For the initial condition, we use the conformal mapping solution of Eq.~(\ref{conformal}) for $\theta(x,z,0)$, along with $\phi(x,z,0)=-\pi/2$.

\begin{figure}
\includegraphics[width=3.375in]{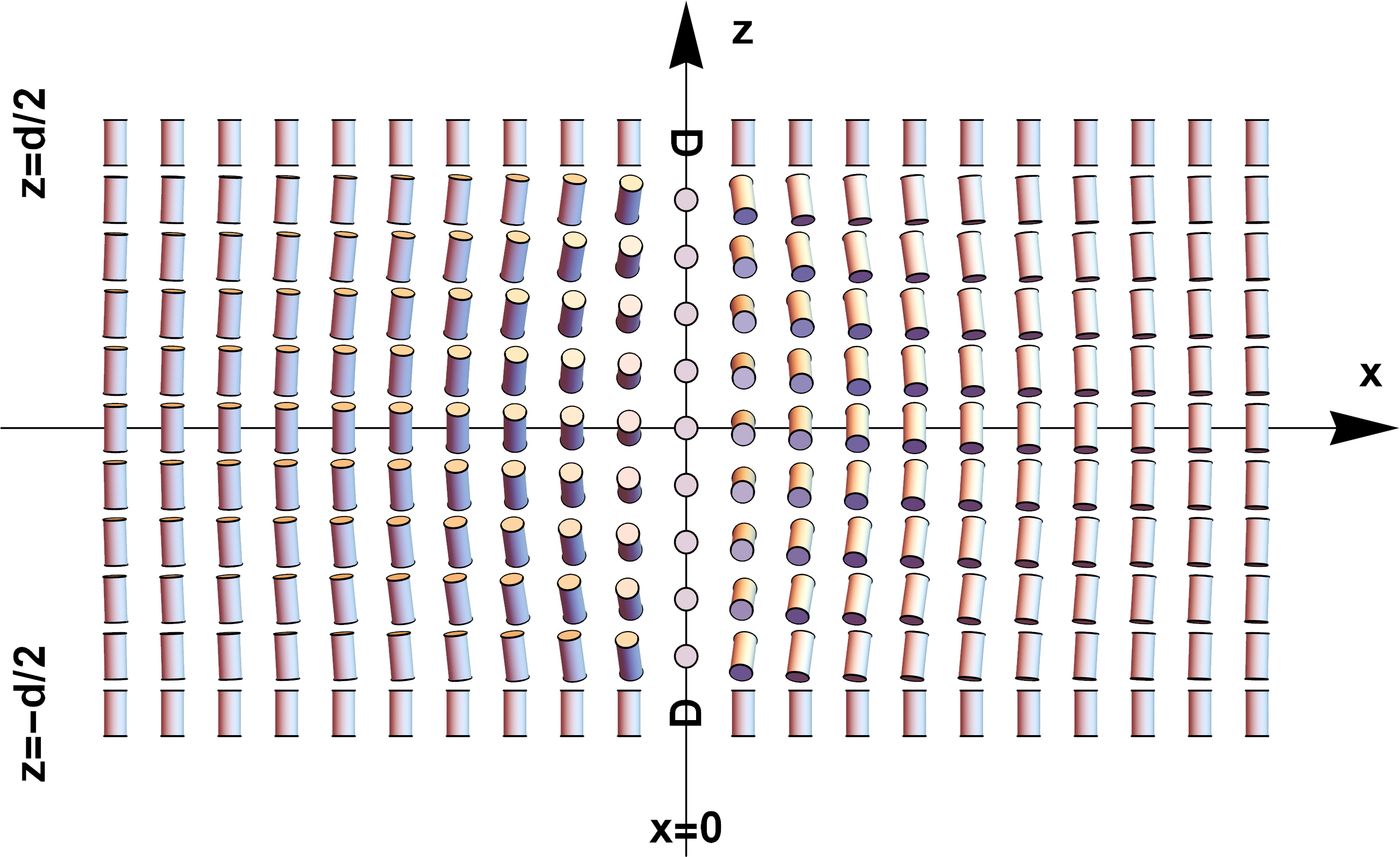}
\caption{(Color online) Single helicoid at $x=0$, calculated numerically without the assumption of constant azimuth.  The picture shows the cross section at $y=0$; the director field is extended uniformly forward and backward in $y$.}
\end{figure}

By integrating the dynamic equations forward in time until they converge, we obtain the director field shown in Fig.~4.  This picture looks generally similar to the conformal mapping result shown in Fig.~1.  However, we can see that the azimuthal angle is not fixed, but rather varies somewhat as a function of both $x$ and $z$.  As a result, the director field has some extra twist from the bottom to the top of the cell.  Because of this extra twist, the free energy of this structure is lower than the free energy found with the assumption of constant azimuth.

To calculate the free energy of the helicoid, we substitute the numerical solution for $\theta$ and $\phi$ into the free energy density of Eq.~(\ref{generalfreeenergy}), and integrate over the whole domain.  In this integration, we have the same problem that was previously discussed in Sec.~II:  the free energy is dominated by the disclinations at top and bottom surfaces, where the integral diverges logarithmically.  To solve this problem, we use the same method as in Sec.~II(A):  we impose a maximum free energy density $f_\text{max}$ as a cutoff on the integrand.  This approach is physically reasonable, because the Frank free energy density can never exceed the free energy density of melting the cholesteric phase into the isotropic phase.  As noted previously, $f_\text{max}$ is related to the disclination core radius $a$ by $f_\text{max}\approx K/a^2$.

\begin{figure}
\includegraphics[width=3.375in]{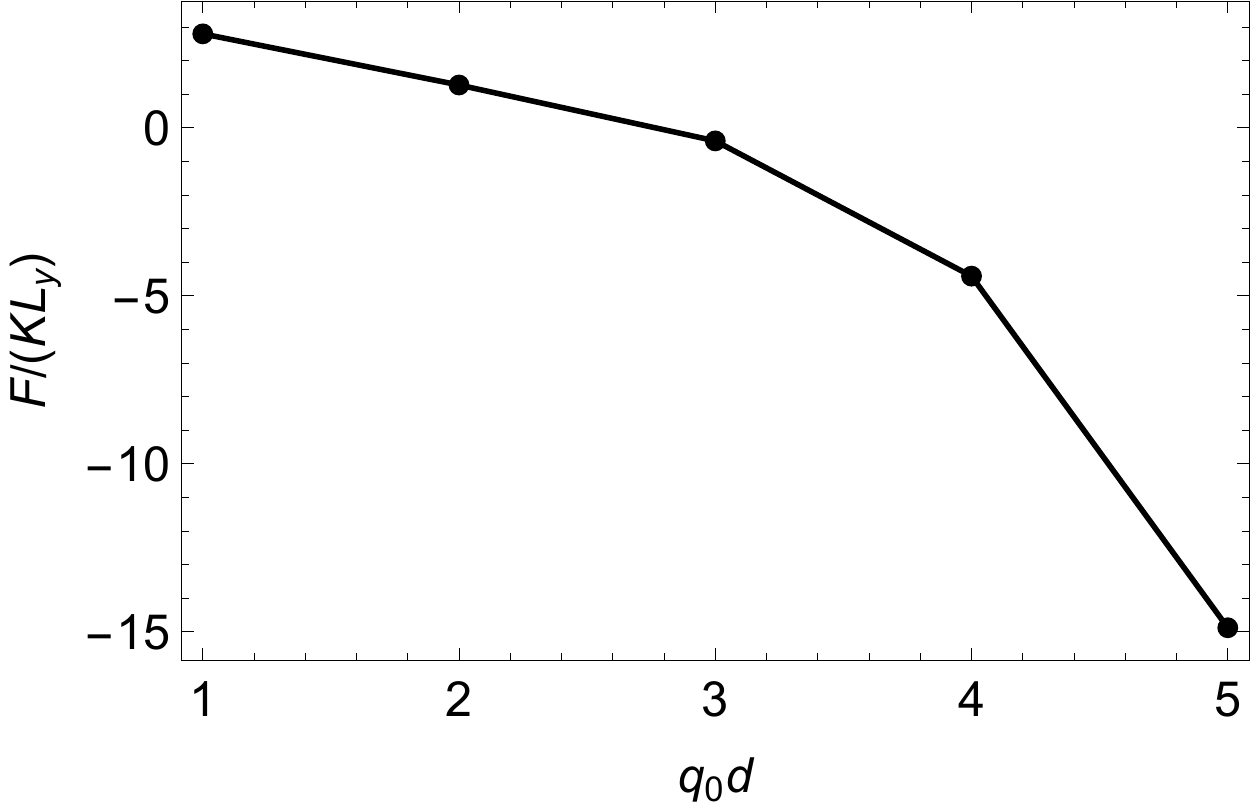}
\caption{Numerical calculation of the free energy of a single helicoid as a function of the natural twist, without the assumption of constant azimuth, using the maximum free energy density $f_\text{max}=100 K/d^2$.}
\end{figure}

Figure~5 shows the numerical result for the integrated free energy of the single helicoid, calculated as a function of the natural twist $q_0$ (scaled by the cell thickness $d$), for fixed $f_\text{max}=100 K/d^2$.  We can see that this numerical result has the general form expected from Eq.~(\ref{fsinglehelicoid}).  There is a critical value $q_H$ where the helicoid free energy crosses from positive to negative.  For $q_0 < q_H$, a helicoid has higher free energy than a uniform vertical alignment (which has $F=0$).  In that case, helicoids will not form in the ground state, although they may occur as metastable defects.  For $q_0 > q_H$, a helicoid has lower free energy than a uniform vertical alignment, and hence helicoids will form in the ground state.

The critical value $q_H$ depends on $f_\text{max}$.  In the example of Fig.~5, we find $q_H \approx 2.8/d$ with the choice $f_\text{max}=100 K/d^2$.  This choice corresponds to $d/a \approx 10$; for example, we might have $d\approx 1$~$\mu$m and $a\approx 100$~nm.  This value of the core radius $a$ is artificially large; it is chosen for numerical convenience, so that the free energy density will not be extremely concentrated in small defect cores.  In experiments, $a$ is normally closer to $10$~nm.  Based on Eq.~(\ref{fsinglehelicoid}), we expect that this reduced value of $a$ would increase $q_H$ by about $(\log 10)/d$, leading to $q_H\approx5/d$.

\subsection{Helicoid lattice}

We now apply the same numerical method to a lattice of parallel helicoids, with a spacing of $\lambda$ in the $x$ direction.  Suppose the centers of the helicoids are located at $x=m\lambda$, where $m$ is any integer.  We solve the dynamic equations in a domain between two helicoids, for $0\le x\le\lambda$ and $-\frac{1}{2}d\le z\le\frac{1}{2}d$.  The rest of the director field can be found by repeating this unit cell periodically.

For the boundary conditions, we require
\begin{align}
&\theta(0,z,t)=\frac{\pi}{2},        &&\phi(0,z,t)=-\frac{\pi}{2},\nonumber\\
&\theta(\lambda,z,t)=\frac{3\pi}{2}, &&\phi(\lambda,z,t)=-\frac{\pi}{2},\nonumber\\
&\theta(x,\frac{d}{2},t)=\pi,        &&\phi(x,\frac{d}{2},t)=-\frac{\pi}{2},\nonumber\\
&\theta(x,-\frac{d}{2},t)=\pi,       &&\phi(x,-\frac{d}{2},t)=-\frac{\pi}{2}.
\end{align}
For the initial condition, we construct a combination of two helicoids using displaced versions of the conformal mapping solution~(\ref{conformal}),
\begin{align}
\theta(x,z,0)= & \pi+\tan^{-1}\left(\frac{\sin(\pi z/d)-\exp(-\pi x/d)}{\cos(\pi z/d)}\right)\\
&-\tan^{-1}\left(\frac{\sin(\pi z/d)+\exp(-\pi x/d)}{\cos(\pi z/d)}\right)\nonumber\\
&-\tan^{-1}\left(\frac{\sin(\pi z/d)-\exp(-\pi [\lambda-x]/d)}{\cos(\pi z/d)}\right)\nonumber\\
&+\tan^{-1}\left(\frac{\sin(\pi z/d)+\exp(-\pi [\lambda-x]/d)}{\cos(\pi z/d)}\right),\nonumber
\end{align}
along with $\phi(x,z,0)=-\pi/2$.

\begin{figure}
\includegraphics[width=3.375in]{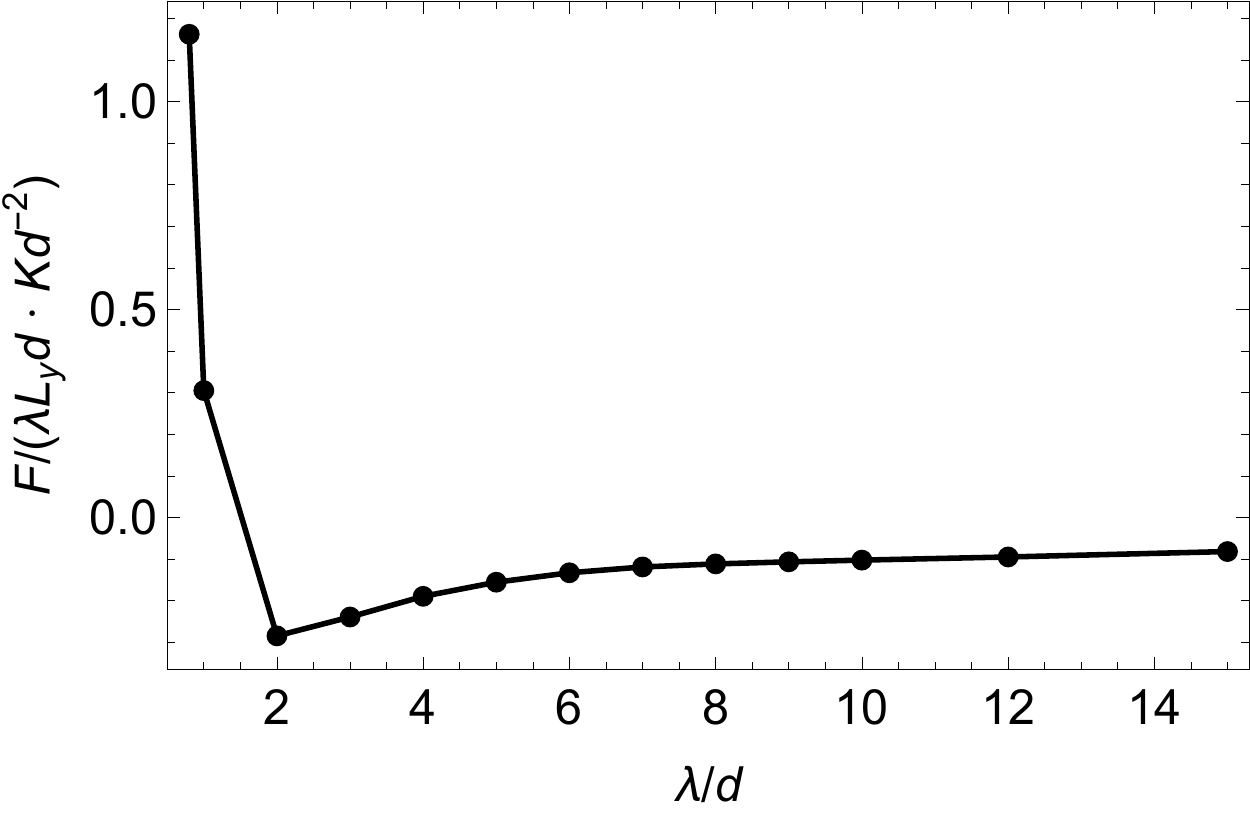}
\caption{Numerical calculation of the average free energy per volume of a helicoid lattice, as a function of the periodicity $\lambda$, using the natural twist $q_0 = 3/d$ and the maximum free energy density $f_\text{max}=100 K/d^2$, without the assumption of constant azimuth.}
\end{figure}

\begin{figure}
\includegraphics[width=3.375in]{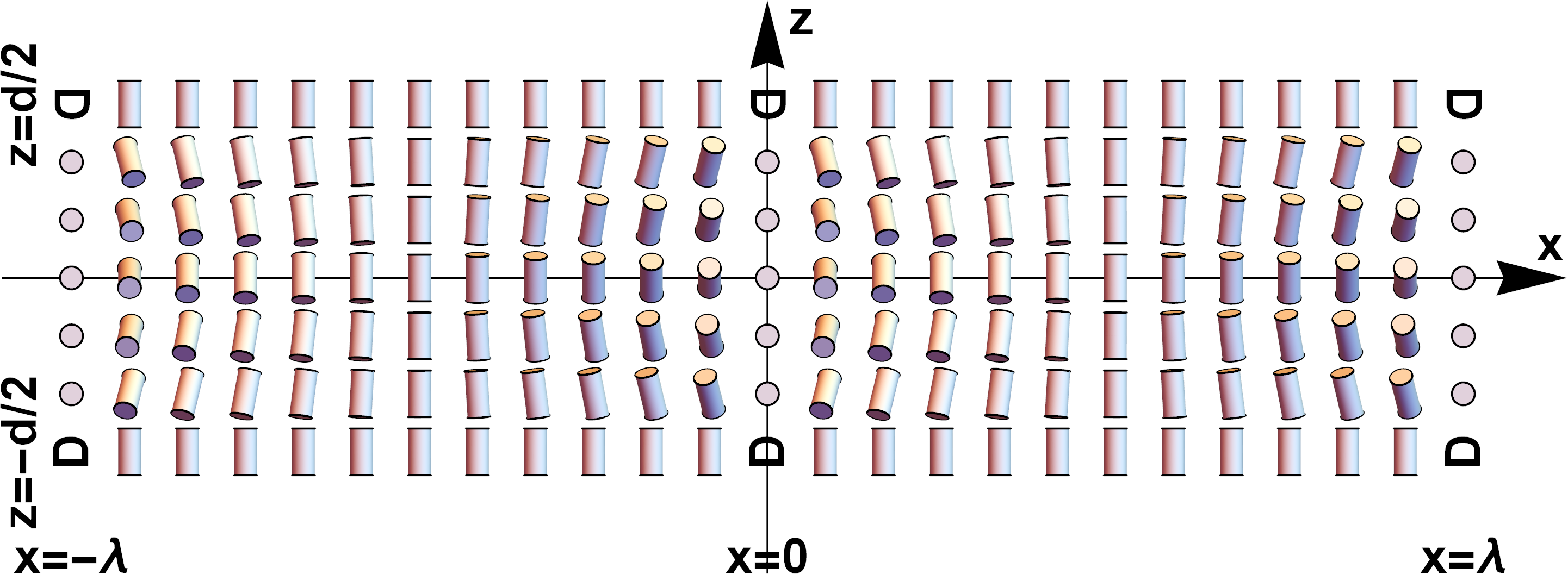}
\caption{(Color online) Lattice of helicoids at $x=-\lambda$, $0$, $\lambda$, \ldots, calculated numerically without the assumption of constant azimuth.  This example is constructed using the natural twist $q_0 = 3/d$ and the maximum free energy density $f_\text{max}=100 K/d^2$, and hence the periodicity is $\lambda\approx 2d$.  The picture shows the cross section at $y=0$; the director field is extended uniformly forward and backward in $y$.}
\end{figure}

By integrating the dynamic equations, we calculate the director field for several values of the periodicity $\lambda$.  We then perform a numerical integration to determine the free energy of the helicoid lattice.  The results depend on $\lambda$ as well as the natural twist $q_0$ and maximum free energy density $f_\text{max}$.  As an example, Fig.~6 shows the free energy per volume $F/(\lambda L_y d)$ (scaled by $K/d^2$), as a function of $\lambda$ (scaled by $d$), for $q_0 = 3/d$ and $f_\text{max}=100 K/d^2$.  For these parameters, the free energy per volume has a minimum at $\lambda\approx 2d$, and hence the helicoid lattice will form with that optimum spacing.  Figure~7 shows the director field that corresponds to this minimum free energy per volume.  It is a periodic sequence of helicoids, separated by regions where the director field is predominantly vertical.  Each helicoid in the lattice has variation in both $\theta$ and $\phi$, similar to the single helicoid shown in Fig.~4.

We now repeat the helicoid lattice calculation for different values of the natural twist $q_0$ and the maximum free energy density $f_\text{max}$.  In each case, we determine the optimum helicoid spacing $\lambda$, as well as the free energy per volume at that spacing.  For any fixed $f_\text{max}$, the dependence of $\lambda$ on $q_0$ is similar to the prediction of Fig.~2.  When $q_0$ is slightly above the critical value $q_H$, the helicoid spacing $\lambda$ is large.  As $q_0$ increases, $\lambda$ decreases.  Over a range of $q_0$, $\lambda$ is close to the cell thickness $d$.  The critical value $q_H$ increases as $f_\text{max}$ increases.

The results for free energy per volume at the optimum helicoid spacing will be used to compare the lattice of helicoids with the lattice of skyrmions discussed in the next section.

\section{Skyrmions: Numerical solutions without assumption of constant azimuth}

Apart from helicoids, another way for a cholesteric liquid crystal to adapt to confinement within a thin cell is to form skyrmions.  While a helicoid is narrow in $x$ and extended in $y$, a skyrmion is narrow in both $x$ and $y$, so that it is a point-like object in the $xy$ plane.  As discussed in Ref.~\cite{Leonov2014}, the director field associated with a skyrmion is vertical in the center, then twists going radially outward, then becomes vertical again far from the center.

Here, as in Ref.~\cite{Leonov2014}, we will assume that a skyrmion is axisymmetric, i.e.\ rotationally symmetric about its central axis.  For that reason, it is most convenient to represent the director field of a skyrmion in cylindrical coordinates $(\rho,\Phi,z)$, so that
\begin{align}
\bm{\hat n}=
&\bm{\hat x}\sin\theta(x,y,z)\cos\phi(x,y,z)\nonumber\\
&+\bm{\hat y}\sin\theta(x,y,z)\sin\phi(x,y,z)+\bm{\hat z}\cos\theta(x,y,z)\nonumber\\
=&\bm{\hat\rho}\sin\theta(\rho,z)\cos\delta\phi(\rho,z)\nonumber\\
&+\bm{\hat\Phi}\sin\theta(\rho,z)\sin\delta\phi(\rho,z)+\bm{\hat z}\cos\theta(\rho,z).
\label{ncylindrical}
\end{align}
Here, $\phi$ is the angle with respect to the $x$ axis, while $\delta\phi=\phi-\Phi$ is the angle with respect to the local radial direction $\bm{\hat\rho}$.  If the skyrmion is axisymmetric, then $\theta$ and $\delta\phi$ can only be functions of $\rho$ and $z$; they must be independent of $\Phi$.

Inserting the director field~(\ref{ncylindrical}) into the Frank free energy density~(\ref{frank}) gives
\begin{align}
f=\frac{1}{2}K\biggl[
&\left(\frac{\partial\theta}{\partial z}\right)^2
+\left(\frac{\partial\theta}{\partial\rho}\right)^2
+\frac{\sin2\theta}{\rho} \frac{\partial\theta}{\partial\rho}
+\frac{\sin^2 \theta}{\rho^2}\nonumber\\
&+\sin^2 \theta \biggl(\left(\frac{\partial\delta\phi}{\partial\rho}\right)^2 + \left(\frac{\partial\delta\phi}{\partial z}\right)^2\biggr)\nonumber\\
&+2\sin^2 \theta \sin\delta\phi \left(\frac{\partial\theta}{\partial z} \frac{\partial\delta\phi}{\partial\rho}
-\frac{\partial\delta\phi}{\partial z} \frac{\partial\theta}{\partial\rho}\right)\nonumber\\
&-\frac{2\sin^2 \theta \cos\delta\phi}{\rho} \frac{\partial\theta}{\partial z}\nonumber\\
&+2q_0 \sin\delta\phi \frac{\partial\theta}{\partial\rho}
-2q_0 \sin^2 \theta \frac{\partial\delta\phi}{\partial z}\nonumber\\
&+q_0 \sin2\theta \cos\delta\phi \frac{\partial\delta\phi}{\partial\rho}
+\frac{q_0 \sin2\theta \sin\delta\phi}{\rho}
\biggr].
\label{generalfreeenergyskyrmion}
\end{align}
The integrated free energy in cylindrical coordinates is
\begin{equation}
F=\int 2\pi\rho d\rho dz f(\rho,z).
\end{equation}
We now have a situation similar to Eq.~(\ref{generalfreeenergy}) for helicoids, but in cylindrical coordinates.  It is not obvious whether the azimuthal angle $\delta\phi$ should be constant with respect to $\rho$ and $z$.  If $\delta\phi$ were constant, its optimal value would be $\delta\phi=-\pi/2$, in order to minimize the term $K q_0 \sin\delta\phi (\partial\theta/\partial\rho)$ in the free energy.  In this case, the director distortion would be mainly twist, rather than splay or bend.  Based on these considerations, Ref.~\cite{Leonov2014} made the assumption that $\delta\phi=-\pi/2$ throughout the cell, and calculated the resulting director configuration around the skyrmion.  However, in Sec.~III, we found that helicoids can reduce their free energy by allowing their azimuthal angle to vary.  Hence, we now apply the same numerical method to determine whether skyrmions can also reduce their free energy by allowing the azimuth to vary.

For this calculation, we set up the dynamic equations
\begin{align}
\frac{\partial\theta(\rho,z,t)}{\partial t}=&-\Gamma_\theta \frac{\delta F}{\delta\theta(\rho,z,t)},\nonumber\\
\frac{\partial(\delta\phi(\rho,z,t))}{\partial t}=&-\Gamma_{\delta\phi} \frac{\delta F}{\delta(\delta\phi(\rho,z,t))}.
\end{align}
We set the dynamic constants $\Gamma_\theta$ and $\Gamma_{\delta\phi}$ equal to each other, and choose units of time so they are $1$.  We use initial conditions and boundary conditions appropriate for the geometry of a single skyrmion or a skyrmion lattice, and integrate the dynamic equations forward in time until they converge on a free energy minimum.

\subsection{Single skyrmion}

To model a single skyrmion at the origin, we solve the dynamic equations in the domain $\rho_\text{min}\le\rho\le\rho_\text{max}$ and $-\frac{1}{2}d\le z\le\frac{1}{2}d$.  Here, $\rho_\text{min}$ is a short-distance cutoff to avoid a singularity in the numerical method at $\rho=0$, and $\rho_\text{max}$ is a long-distance cutoff where the director field is effectively vertical.  We use $\rho_\text{min} = 0.001d$ and $\rho_\text{max} = 10d$.  The appropriate boundary conditions are
\begin{align}
&\theta(\rho_\text{min},z,t)=0,   &&\delta\phi(\rho_\text{min},z,t)=-\frac{\pi}{2},\nonumber\\
&\theta(\rho_\text{max},z,t)=\pi, &&\delta\phi(\rho_\text{max},z,t)=-\frac{\pi}{2},\nonumber\\
&\theta(\rho,\frac{d}{2},t)=\pi,  &&\delta\phi(\rho,\frac{d}{2},t)=-\frac{\pi}{2},\nonumber\\
&\theta(\rho,-\frac{d}{2},t)=\pi, &&\delta\phi(\rho,-\frac{d}{2},t)=-\frac{\pi}{2}.
\label{skyrmionbc}
\end{align}
The boundary conditions on $\delta\phi$ are not important, because the boundary conditions on $\theta$ require that the director field is vertical on all the boundaries, but we include them as part of the numerical algorithm.

For the initial condition, we use a modified version of the conformal mapping solution from Eq.~(\ref{conformal}),
\begin{align}
\theta(\rho,z,0)= & \pi+2\tan^{-1}\left(\frac{\sin(\pi z/d)-\exp(-\pi\rho/(2d))}{\cos(\pi z/d)}\right)\nonumber\\
&-2\tan^{-1}\left(\frac{\sin(\pi z/d)+\exp(-\pi\rho/(2d))}{\cos(\pi z/d)}\right),
\end{align}
along with $\delta\phi(\rho,z,0)=-\pi/2$.  We recognize that this expression for $\theta(\rho,z,0)$ is not the exact solution of any problem in cylindrical coordinates, but it has the correct topological form for a skyrmion and is useful as a starting point for the numerical algorithm.

\begin{figure}
\includegraphics[width=3.375in]{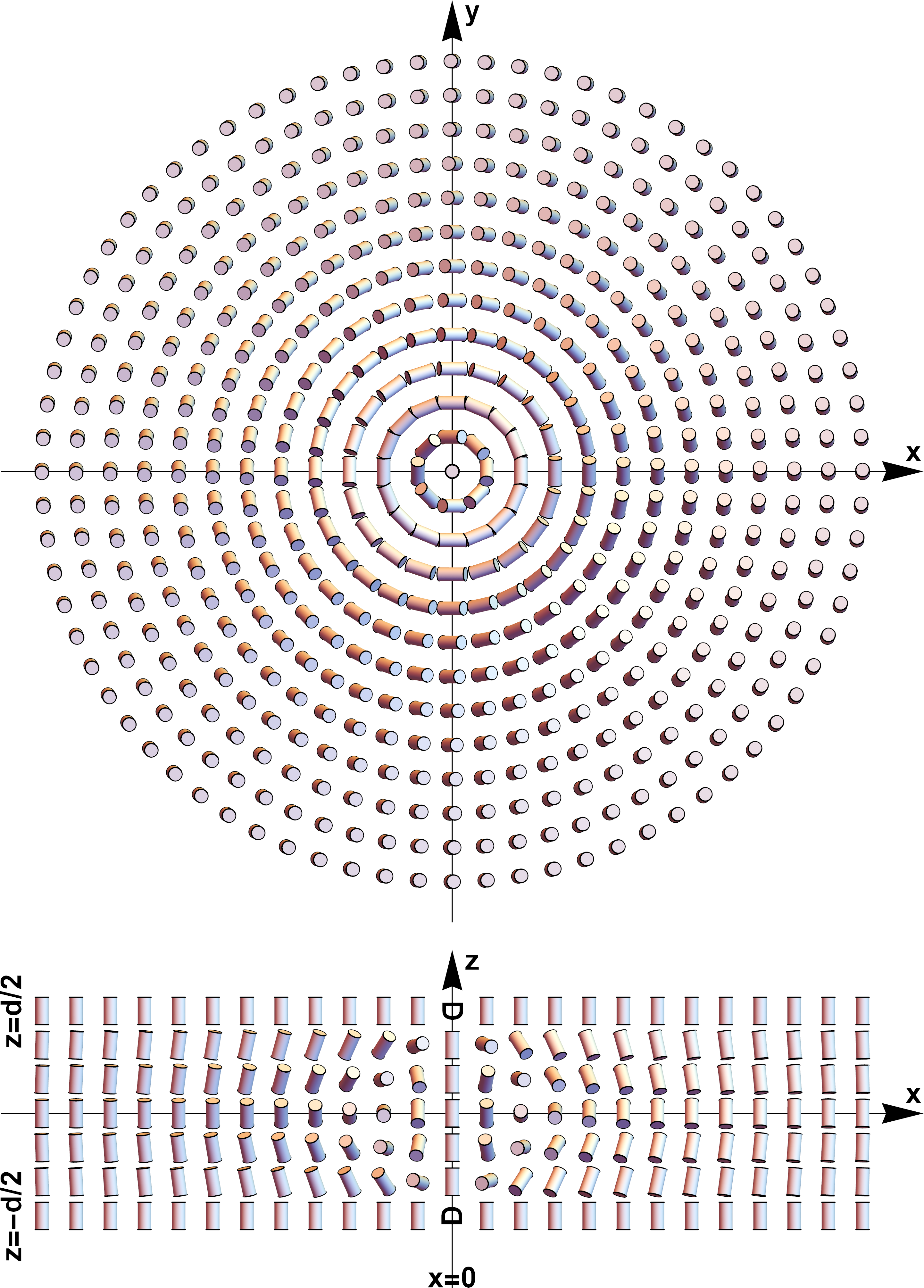}
\caption{(Color online) Director field of a single skyrmion, calculated numerically for natural twist $q_0 = 3/d$.  Images show the horizontal cross section at $z=0$ and the vertical cross section at $y=0$.  The symbol D represents point defects on the top and bottom surfaces.}
\end{figure}

We integrate the dynamic equations until they converge on the director field of a single skyrmion.  Figure~8 shows an example, calculated for natural twist $q_0 = 3/d$, with both a horizontal cross section at $z=0$ and a vertical cross section at $y=0$.  In this configuration, the director field twists radially outward from the central axis, and it \emph{also} twists from the bottom to the top of the cell.  The director field has point defects where the central axis intersects the top and bottom surfaces, at $\rho=0$ and $z=\pm d/2$.  These point defects are exceptions to the general rule that the magnitude of nematic order is constant everywhere in a skyrmion.

To calculate the free energy of the skyrmion, we substitute the numerical solution for $\theta$ and $\delta\phi$ into the free energy density of Eq.~(\ref{generalfreeenergyskyrmion}), and integrate over the whole domain.  Although the free energy density diverges at the point defects on the top and bottom surfaces, we can integrate over these divergences because they are only points in 3D.  For that reason, the total free energy of the skyrmion is not sensitive to the maximum free energy density $f_\text{max}$; it has a well-behaved limit as $f_\text{max}\to\infty$.  We verify numerically that the total free energy does not depend significantly on $f_\text{max}$ for $f_\text{max} > 100 K/d^2$.  Hence, we only report the skyrmion free energy in the limit of large $f_\text{max}$.

\begin{figure}
\includegraphics[width=3.375in]{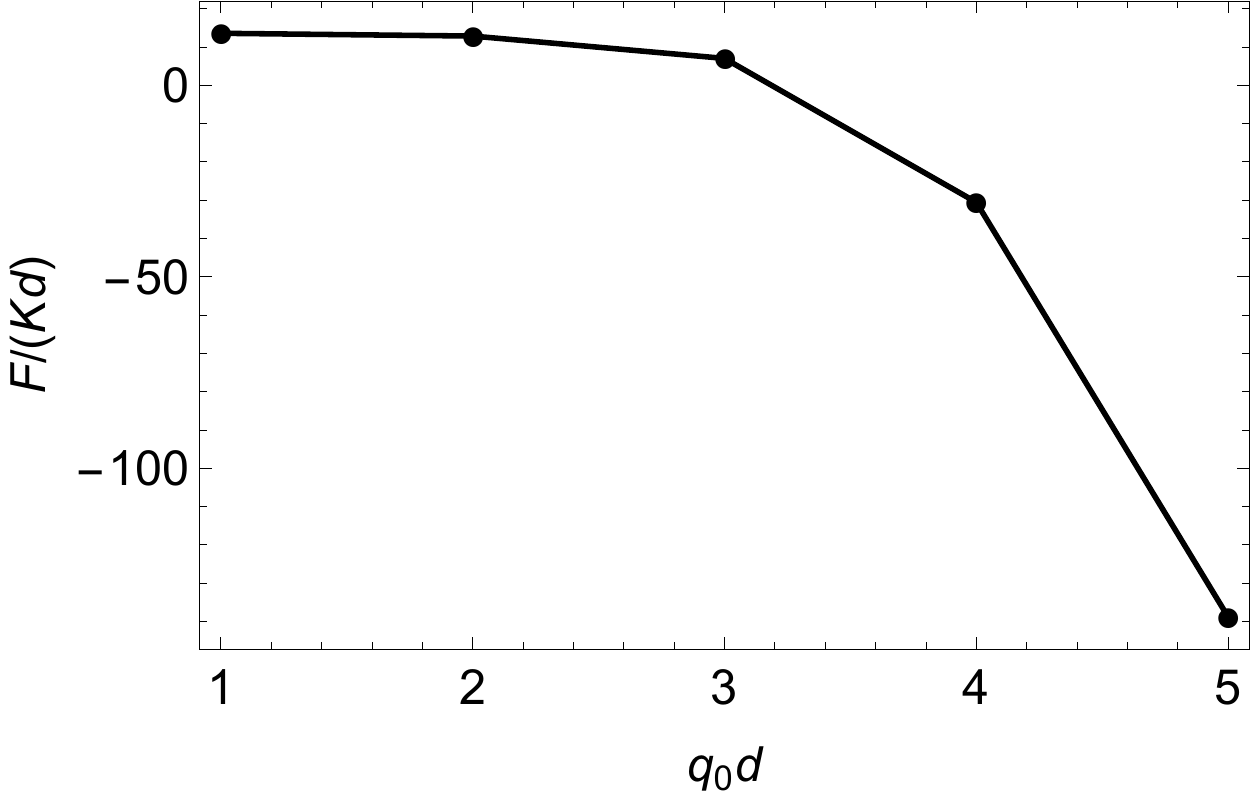}
\caption{Free energy of a single skyrmion as a function of natural twist $q_0$.}
\end{figure}

Figure 9 shows the integrated free energy of the single skyrmion as a function of the natural twist $q_0$, scaled by the cell thickness $d$.  This plot is similar to Fig.~5 for a single helicoid.  Here, the skyrmion free energy crosses from positive to negative at a critical value $q_S \approx 3.3/d$.  For $q_0 < q_S$, a skyrmion has higher free energy than a uniform vertical alignment (with $F=0$), and hence skyrmions will not form in thermal equilibrium, although they may occur as metastable defects.  For $q_0 > q_S$, a skyrmion has lower free energy than a uniform vertical alignment, and hence many skyrmions will be present in thermal equilibrium.  To find the favored density of skyrmions, we must consider a periodic lattice of skyrmions below.

\subsection{Skyrmion lattice}

\begin{figure}
\includegraphics[width=3.375in]{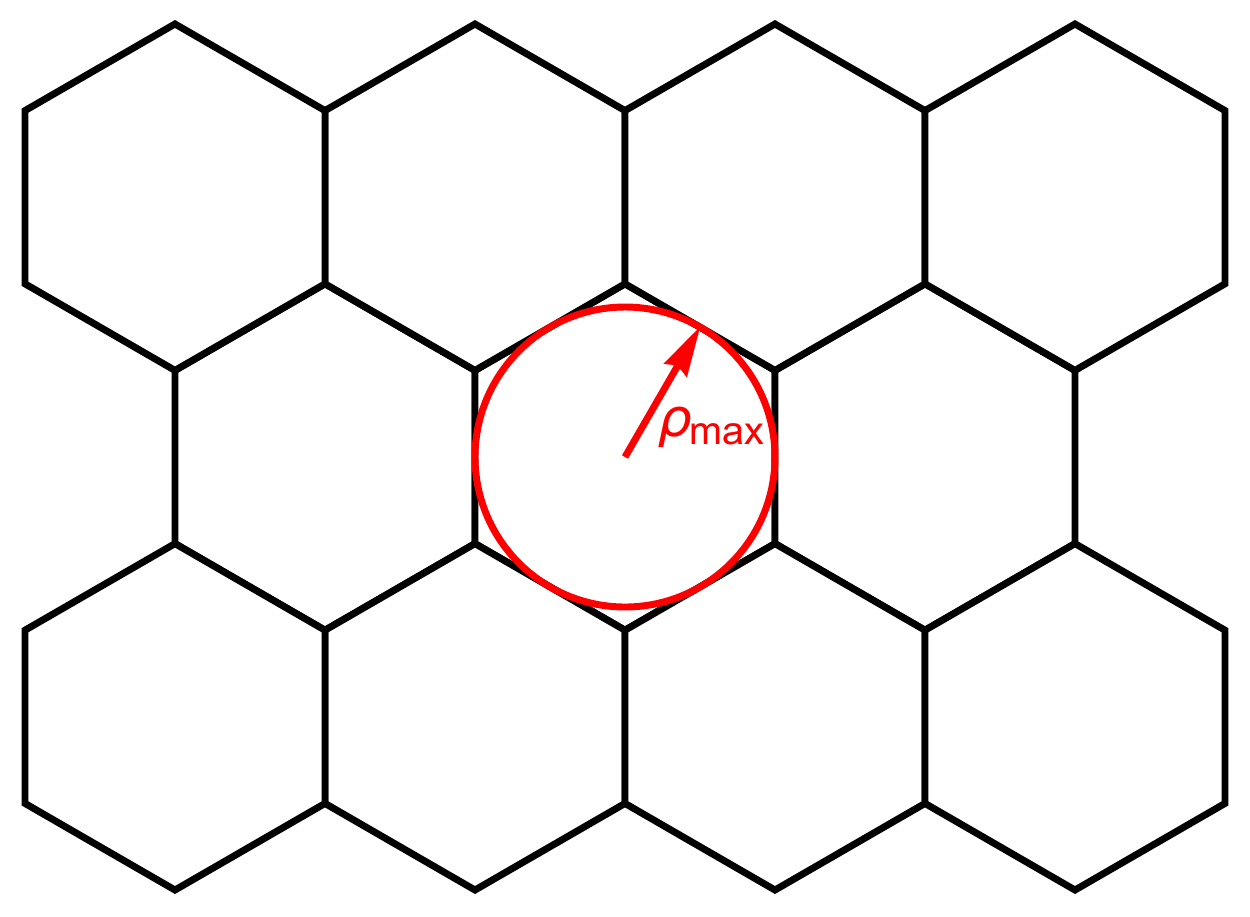}
\caption{(Color online) Hexagonal lattice of skyrmions.  The hexagonal unit cell is approximated by a circle of radius $\rho_\text{max}$.}
\end{figure}

Suppose that a liquid crystal has a periodic lattice of skyrmions.  A two-dimensional lattice of point-like objects normally has a hexagonal structure, as shown in Fig.~10.  To describe this lattice, we should calculate the director configuration within a hexagonal unit cell, and integrate the free energy density over the unit cell.  This calculation is difficult because the hexagon is not exactly axisymmetric, and hence the director field and free energy density depend slightly on the angular coordinate $\Phi$ as well as on $\rho$ and $z$.

To avoid this difficulty, we use the \emph{circular cell approximation}, as is done in Ref.~\cite{Leonov2014} and in many papers on magnetic skyrmions.  We  approximate the hexagonal unit cell by a circle of radius $\rho_\text{max}$, as shown in red in Fig.~10.  We then have the much simpler problem of calculating the director field and integrating the free energy density in the axisymmetric circular cell, where the director field and free energy density depend only on $\rho$ and $z$.

Following this approximation, we solve the dynamic equations in the domain $\rho_\text{min}\le\rho\le\rho_\text{max}$ and $-\frac{1}{2}d\le z\le\frac{1}{2}d$, with the boundary conditions of Eq.~(\ref{skyrmionbc}).  Evidently, this is the same numerical problem that we solved for a single skyrmion in Sec.~IV(A).  The only difference is in the interpretation of the radius $\rho_\text{max}$.  In Sec.~IV(A), we considered the limit of very large $\rho_\text{max}$, much greater than the cell thickness $d$, and we calculated the free energy of a single skyrmion in an effectively infinite domain.  Here, we consider $\rho_\text{max}$ as a lattice spacing, comparable to $d$, and calculate the free energy per volume of each cell in the skyrmion lattice.

\begin{figure}
\includegraphics[width=3.375in]{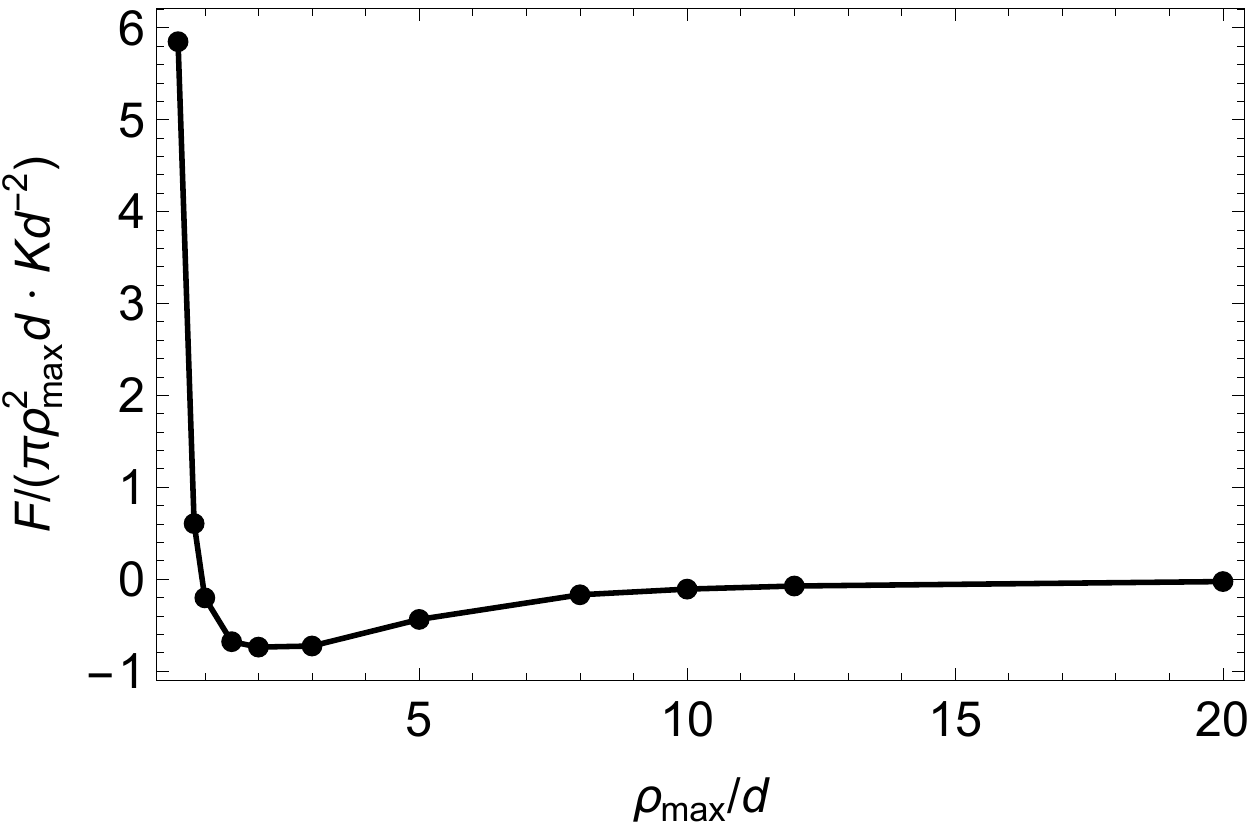}
\caption{Average free energy per unit volume of the skyrmion lattice $F/(\pi\rho_\text{max}^2 d)$ (scaled by $K/d^2$), as a function of the unit cell radius $\rho_\text{max}$ (scaled by $d$), using the natural twist $q_0 = 4/d$.}
\end{figure}

By integrating the dynamic equations, we determine the director field for several values of $\rho_\text{max}$, and then perform a numerical integration to find the free energy of each unit cell.  The results depend on $\rho_\text{max}$ as well as the natural twist $q_0$.  As an example, Fig.~11 shows the free energy per volume $F/(\pi\rho_\text{max}^2 d)$ (scaled by $K/d^2$), as a function of $\rho_\text{max}$ (scaled by $d$), for $q_0 = 4/d$.  For this natural twist, the free energy per volume has a minimum at $\rho_\text{max}=2d$, and hence the skyrmion lattice will form with that optimum unit cell radius.

We now repeat the skyrmion lattice calculation for different values of the natural twist $q_0$.  For each $q_0$, we determine the optimum cell radius $\rho_\text{max}$, as well as the free energy per volume at that radius.  In the next section, the free energy results will be used to compare the lattice of skyrmions with the lattice of helicoids.

\section{Phase diagram}

In the previous two sections, we calculated the average free energy per volume for two topological structures, the helicoid lattice and the skyrmion lattice.  These free energies are both calculated with respect to the uniform vertical configuration, which has $F=0$, and they are both scaled by the same factor $K/d^2$.  Hence, we can compare them to determine which structure is favored:  the helicoid lattice, the skyrmion lattice, or the uniform vertical configuration.

\begin{figure}
\includegraphics[width=3.375in]{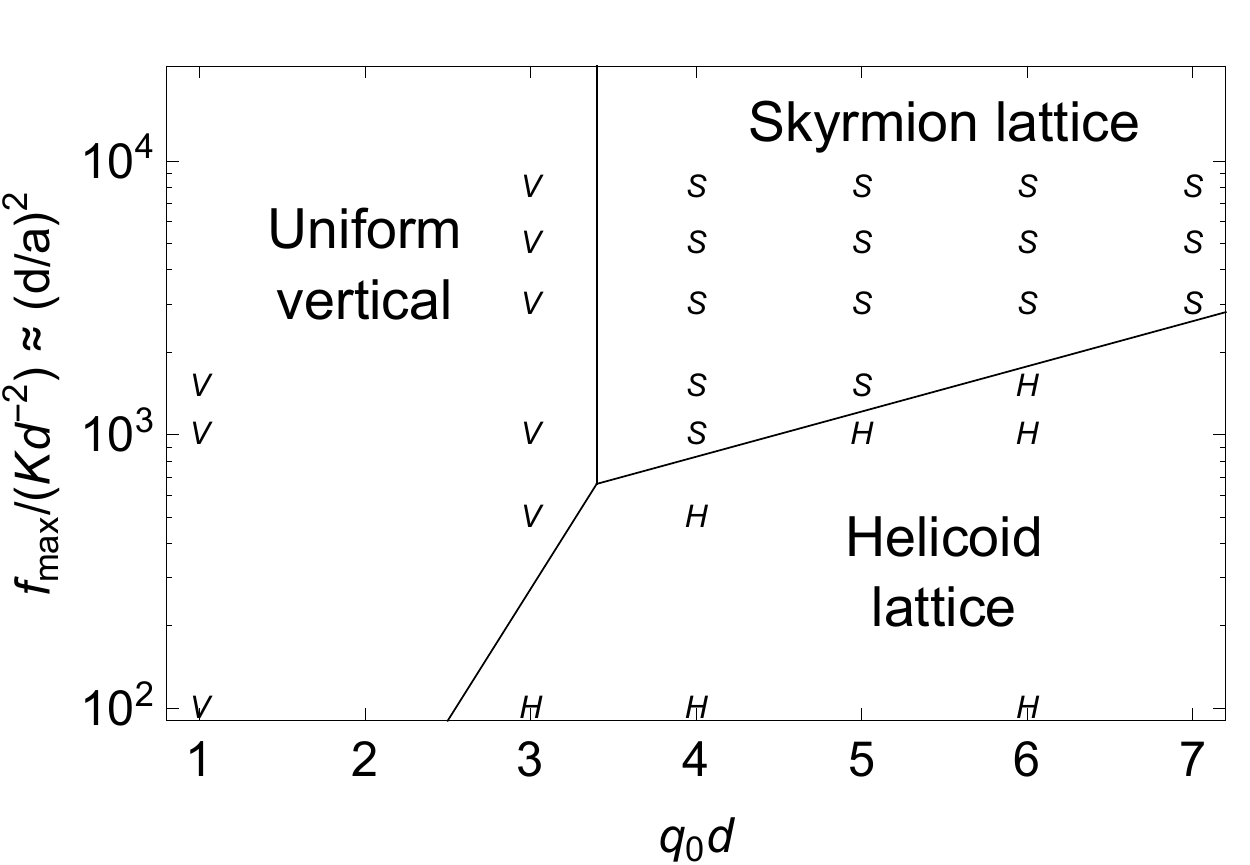}
\caption{Phase diagram indicating the uniform vertical, skyrmion lattice, and helicoid lattice phases, as functions of the natural twist $q_0$ and maximum free energy density $f_\text{max}$.  The symbols $V$, $S$, and $H$ indicate numerical calculations of the lowest free energy structure, while the lines are guides to the eye.  As discussed in the text, the vertical axis can also be interpreted as the ratio of the cell thickness $d$ to the disclination core radius $a$.}
\end{figure}

The phase diagram of Fig.~12 shows the favored structure as a function of two parameters:  the natural twist $q_0$ (scaled by cell thickness $d$) and the maximum free energy density $f_\text{max}$ (scaled by $K/d^2$).  When $f_\text{max}$ is small, there is a direct transition from the uniform vertical configuration to the helicoid lattice as $q_0$ increases.  For larger $f_\text{max}$, there is one transition from uniform vertical to skyrmion lattice, and then another transition from skyrmion lattice to helicoid lattice.  The value of $q_0$ needed to obtain the helicoid lattice increases as $f_\text{max}$ increases.

The structure of this phase diagram can be understood intuitively through the following argument.  The advantage of the helicoid lattice compared with the skyrmion lattice is that it has a lower \emph{bulk} free energy.  The helicoid lattice is closer to the perfect cholesteric helix, which is the equilibrium phase in the bulk.  By comparison, the advantage of the skyrmion lattice compared with the helicoid lattice is that it has a lower \emph{surface} free energy.  The skyrmion lattice has only point defects in the director field on the surface, while the helicoid lattice has disclination lines running along the surface, with an energy per unit length that is proportional to $\log(f_\text{max}d^2 /K)$.  Hence, the helicoid lattice is favored for large $q_0$ and small $f_\text{max}$ (where the bulk free energy dominates), while the skyrmion lattice is favored for large $f_\text{max}$ and smaller $q_0$ (where the surface free energy dominates).  When $q_0$ becomes even smaller, compared with $d$, the benefit from the chiral terms in the free energy becomes smaller than the cost of director gradients from the nonchiral terms, and the uniform vertical configuration is favored over either type of chiral lattice.

Although we have only done calculations for the limit of \emph{infinitely strong} homeotropic anchoring on the surfaces, we can anticipate what would happen if the homeotropic anchoring had only a finite strength $W$ per unit area.  If $W$ were reduced, it would be easier to form line defects on the surface, and hence it would be easier to form the helicoid lattice.  As a result, the helicoid lattice would occur for a lower value of $q_0$.  In Fig.~13, the vertical axis should really be interpreted as the free energy cost of surface defects.  That free energy cost is controlled by $\log(f_\text{max}d^2 /K)$ or $\log(W d/K)$, whichever is smaller.  In this paper, we have done calculations for $W\to\infty$, so that $\log(f_\text{max}d^2 /K)$ is the relevant scale.  For weaker anchoring, $\log(W d/K)$ might become the relevant scale instead.  (In terms of the disclination core radius $a\approx(f_\text{max}/K)^{-1/2}$ and the surface extrapolation length $b\approx K/W$, the scale is $\log(d/a)$ or $\log(d/b)$, whichever is smaller.  Hence, the relevant length is $a$ or $b$, whichever is larger.  In this paper, we have done calculations for $b\to0$, so that $a$ is the relevant length.  For weaker anchoring, $b$ might become the relevant length.)

We should point out one peculiar discrepancy between our results and Leonov et al~\cite{Leonov2014}.  In our phase diagram, the sequence of structures is uniform vertical, then skyrmion lattice, then helicoid lattice.  By contrast, in their phase diagram, the sequence of structures is ``isolated skyrmions'' (i.e.\ uniform vertical with metastable skyrmions), then helicoid lattice, then skyrmion lattice.  Of course, their system is somewhat different from our system, because they have an applied electric field and we do not.  Even so, it is surprising that the sequence of structures would be different.  Resolving this discrepancy should be a subject for future theoretical research, as well as comparison with experiment.

In conclusion, we have developed a theory for helicoids and skyrmions in cholesteric liquid crystals that are confined between surfaces with homeotropic anchoring.  Our work extends previous theoretical research by deriving exact solutions for helicoids with constant azimuth, by calculating numerical solutions for helicoids and skyrmions with varying azimuth, and by interpreting the results in terms of competition between different terms in the free energy.  It provides specific examples of the general principle that complex topological structures are induced by geometric frustration, as seen in chiral magnets as well as liquid crystals.

\acknowledgments

We would like to thank A. Duzgun, A. Saxena, R. L. B. Selinger, and Q. Wei for helpful discussions.  This work was supported by National Science Foundation Grant No.~DMR-1409658.

\bibliography{skyrmion}

\end{document}